\documentclass[onecolumn,english,conference,10pt,twocolumn]{IEEEtran}
\usepackage[T1]{fontenc}
\usepackage[utf8]{inputenc}
\usepackage{xcolor}
\usepackage{float}
\usepackage{mathrsfs}
\usepackage{amsmath}
\usepackage{amsthm}
\usepackage{amssymb}
\usepackage{graphicx}
\usepackage{geometry}
\PassOptionsToPackage{normalem}{ulem}
\usepackage{ulem}

\makeatletter

\floatstyle{ruled}
\newfloat{algorithm}{tbp}{loa}
\providecommand{\algorithmname}{Algorithm}
\floatname{algorithm}{\protect\algorithmname}
\providecolor{lyxadded}{rgb}{0,0,1}
\providecolor{lyxdeleted}{rgb}{1,0,0}
\DeclareRobustCommand{\mklyxadded}[1]{\textcolor{lyxadded}\bgroup#1\egroup}
\DeclareRobustCommand{\mklyxdeleted}[1]{\textcolor{lyxdeleted}\bgroup\mklyxsout{#1}\egroup}
\DeclareRobustCommand{\mklyxsout}[1]{\ifx\\#1\else\sout{#1}\fi}

\DeclareRobustCommand{\lyxdeleted}[4][]{\mklyxdeleted{#4}}

\theoremstyle{plain}
\newtheorem{thm}{\protect\theoremname}
\theoremstyle{remark}

\theoremstyle{definition}
\newtheorem{defn}[thm]{\protect\definitionname}
\theoremstyle{plain}
\newtheorem{lem}[thm]{\protect\lemmaname}
\theoremstyle{plain}
\newtheorem{prop}[thm]{\protect\propositionname}
\theoremstyle{plain}
\newtheorem{cor}[thm]{\protect\corollaryname}

\IEEEoverridecommandlockouts

\usepackage{amsmath,amssymb,amsfonts}
\usepackage[level=0]{wgroup_message}  
\usepackage{graphicx,psfrag,cite,subfigure}

\usepackage{algorithmic}
\usepackage{textcomp}
\usepackage[table]{xcolor}

\author{

\IEEEauthorblockN{Bowen~Li\IEEEauthorrefmark{1}, Jiping~Luo\IEEEauthorrefmark{1}, Themistoklis~Charalambous\IEEEauthorrefmark{2}, Nikolaos~Pappas\IEEEauthorrefmark{1}}
\IEEEauthorblockA{\IEEEauthorrefmark{1} Department of Computer and Information Science, Link{\"o}ping University, 58183, Link{\"o}ping, Sweden}
\IEEEauthorblockA{\IEEEauthorrefmark{2} Department of Electrical
and Computer Engineering, School of Engineering, University of
Cyprus, 1678 Nicosia, Cyprus}
\thanks{This work was supported in part by the European Union (6G-LEADER) under 101192080, and ELLIIT.}
}

\usepackage[acronym]{glossaries}
\newcommand{\newac}{\newacronym}
\newcommand{\ac}{\gls}

\newcommand{\acpl}{\glspl}

\makeglossaries
\newac{speb}{SPEB}{square position error bound}
\newac[plural=EFIMs,firstplural=Fisher information matrices (EFIMs)]{efim}{EFIM}{Fisher information matrix}
\newac{ne}{NE}{Nash equilibrium}
\newac{mse}{MSE}{mean squared error}
\newac{toa}{TOA}{time-of-arrival}
\newac{snr}{SNR}{signal-to-noise ratio}
\newac{lan}{LAN}{local area network}
\newac{psd}{PSD}{positive semidefinite}
\newac{pd}{PD}{positive definite}
\newac{wrt}{w.r.t.}{with respect to}
\newac{lhs}{L.H.S.}{left hand side}
\newac{wp1}{w.p.1}{with probability 1}
\newac{kkt}{KKT}{Karush-Kuhn-Tucker}
\newac{wlog}{w.l.o.g.}{without loss of generality}
\newac{mle}{MLE}{maximum likelihood estimation}
\newac{gps}{GPS}{global positioning system}
\newac{rssi}{RSSI}{received signal strength indicator}
\newac{mimo}{MIMO}{multiple-input multiple-output}
\newac{csi}{CSI}{channel state information}
\newac{fdd}{FDD}{frequency division duplexing}
\newac{ms}{MS}{mobile station}
\newac{bs}{BS}{base station}
\newac{d2d}{D2D}{device-to-device}
\newac{slnr}{SLNR}{signal-to-interference-leakage-and-noise-ratio}
\newac{ula}{ULA}{uniform linear antenna array}
\newac{pas}{PAS}{power angular spectrum}
\newac{mmse}{MMSE}{minimum mean square error}
\newac{zf}{ZF}{zero-forcing}
\newac{rzf}{RZF}{regularized zero-forcing}
\newac{as}{AS}{angular spread}
\newac{aod}{AOD}{angle of departure}
\newac{iid}{i.i.d.}{independent and identically distributed} 
\newac{sinr}{SINR}{signal-to-interference-and-noise ratio}
\newac{tdd}{TDD}{time-division duplex}
\newac{rvq}{RVQ}{random vector quantization}
\newac{rhs}{R.H.S.}{right hand side}
\newac{mrc}{MRC}{maximum ratio combining}
\newac{cdf}{CDF}{cumulative distribution function}
\newac{a.s.}{a.s.}{almost surely}
\newac{los}{LOS}{line-of-sight}
\newac{jsdm}{JSDM}{joint spatial division and multiplexing}
\newac{map}{MAP}{maximum a posteriori}
\newac{klt}{KLT}{Karhunen-Lo\`eve Transform}
\newac{lbe}{LBE}{link bargaining equilibrium}
\newac{se}{SE}{Stackelberg equilibrium}
\newac{uav}{UAV}{unmanned aerial vehicle}
\newac{nlos}{NLOS}{non-line-of-sight}
\newac{pdf}{PDF}{probability density function}
\newac{em}{EM}{expectation-maximization}
\newac{knn}{KNN}{$k$-nearest neighbor}
\newac{svd}{SVD}{singular value decomposition}
\newac{nmf}{NMF}{non-negative matrix factorization}
\newac{umf}{UMF}{unimodality-constrained matrix factorization}
\newac{rmse}{RMSE}{rooted mean squared error}
\newac{olos}{OLOS}{obstructed line-of-sight}
\newac{mmw}{mmW}{millimeter wave}
\newac{ber}{BER}{bit error rate}
\newac{rss}{RSS}{received signal strength}
\newac{lp}{LP}{linear program}
\newac{ufw}{U-FW}{unimodal Frank-Wolfe}
\newac{utf}{UTF}{unimodality-constrained tensor factorization}
\newac{fw}{FW}{Frank-Wolfe}
\newac{iot}{IoT}{Internet-of-Things}
\newac{mae}{MAE}{mean absolute error}
\newac{crb}{CRB}{Cram\'er-Rao bound}
\newac{aoa}{AoA}{angle of arrival}
\newac{wcl}{WCL}{weighted centroid localization}

\newac{mdp}{MDP}{Markov decision process}
\newac{wsn}{WSN}{wireless sensor network}
\newac{iff}{iff}{if and only if}
\newac{qsi}{QSI}{queue state information}
\newac{umi}{UMi}{Urban Micro}
\newac{qos}{QoS}{quality of service}
\newac{aoi}{AoI}{age of information}
\newac{ofdm}{OFDM}{orthogonal frequency division multiplexing}
\newac{rb}{RB}{resource block}
\newac{noma}{NOMA}{non-orthogonal multiple access}

\renewcommand{\lyxdeleted}[3]{{\color{lyxdeleted}{}}}

\definecolor{lightcyan}{rgb}{0.88, 1.0, 1.0}

\makeatother

\usepackage{babel}
\providecommand{\corollaryname}{Corollary}
\providecommand{\definitionname}{Definition}
\providecommand{\lemmaname}{Lemma}
\providecommand{\propositionname}{Proposition}
\providecommand{\remarkname}{Remark}
\providecommand{\theoremname}{Theorem}

\begin{document}
\title{Pareto-Optimal Sampling and Resource Allocation for Timely Communication in Shared-Spectrum Low-Altitude Networks}
\maketitle
\begin{abstract}
Guaranteeing stringent data freshness for low-altitude \acpl{uav} in shared spectrum forces a critical trade-off between two operational costs: the \ac{uav}'s own energy consumption and the occupation of terrestrial channel resources. The core challenge is to satisfy the aerial data freshness while finding a Pareto-optimal balance between these costs. Leveraging predictive channel models and predictive \ac{uav} trajectories, we formulate a bi-objective Pareto optimization problem over a long-term planning horizon to jointly optimize the sampling timing for aerial traffic and the power and spectrum allocation for fair coexistence. However, the problem's non-convex, mixed-integer nature renders classical methods incapable of fully characterizing the complete Pareto frontier. Notably, we show monotonicity properties of the frontier, building on which we transform the bi-objective problem into several single-objective problems. We then propose a new graph-based algorithm and prove that it can find the complete set of Pareto optima with low complexity, linear in the horizon and near-quadratic in the \ac{rb} budget. Numerical comparisons show that our approach meets the stringent timeliness requirement and achieves a six-fold reduction in \ac{rb} utilization or a 6 dB energy saving compared to benchmarks. 
\end{abstract}


\glsresetall

\section{Introduction\label{sec:intro}}
Low-altitude activities have grown significantly over the past decade, resulting in a surge in demand for communication links that support time-critical applications, such as real-time navigation, control, and surveillance \cite{WuXuZenNg:J21}. The core requirement for these services is not merely reliability, but stringent data freshness. For instance, the value of a \ac{uav}'s field of view for mission monitoring is directly tied to its timeliness; a network that fails to deliver this information promptly renders it obsolete, severely compromising situational awareness and operational safety.

\begin{figure}
\begin{centering}
\includegraphics[width=1\columnwidth]{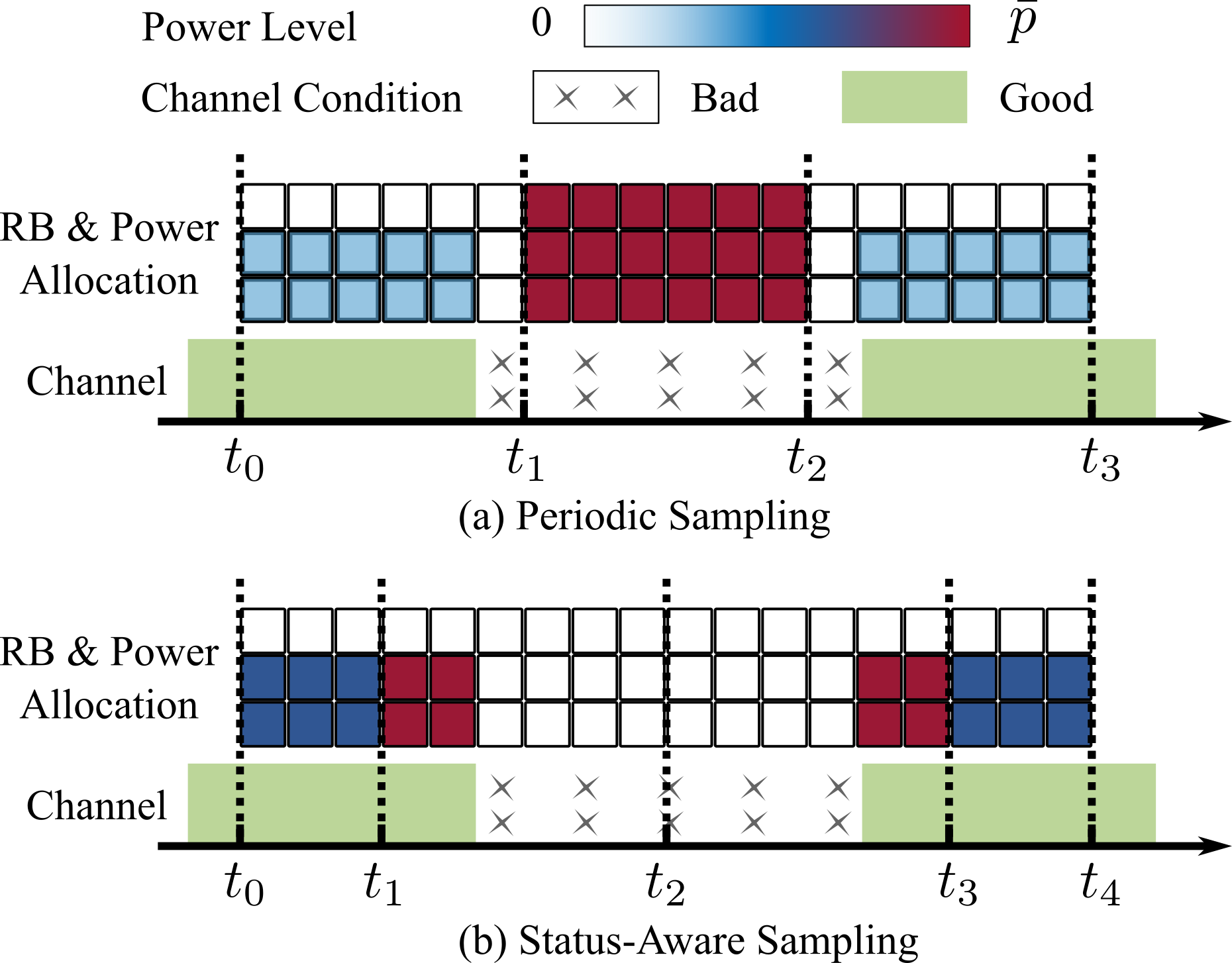}
\par\end{centering}
\caption{\label{fig:status_aware_sampling}Illustration of periodic versus status-aware sampling policies.
(a) A periodic policy enforces equal spacing and causes the update
to fall within a poor-channel interval. (b) A status-aware policy
adapts the sampling time to bypass the poor interval and satisfy the
freshness constraint. Note that although status-aware sampling may
involve more sampling and transmission instants, it exploits better
channel intervals and therefore uses fewer \acpl{rb} and less energy.}
\end{figure}

Most existing works on aerial communications focus on the throughput-reliability-delay trade-off \cite{MatShe:J25,ZhaLiRonZen:J25}. However, such approaches are insufficient to guarantee the stringent timeliness required by low-altitude networks. To address data freshness, some works employ \ac{aoi} and its variants to assess the importance of information and timing and to control sampling and transmission decisions, thereby balancing efficiency and timeliness \cite{AntNikVan:B17, GuoLinLiuKon:J25, LuoPap:J25b, LuoPap:J25, luo2026exploiting, SalKouPap:J24, luo2025information}.
Taking Figure~\ref{fig:status_aware_sampling} as an example, status-aware sampling control can simultaneously reduce power consumption and \ac{rb} allocation while guaranteeing timeliness.
However, the high mobility of \acpl{uav} creates a time-varying topology with rapidly changing channel conditions for both communication and interference links. This highly dynamic environment presents a significant challenge in jointly optimizing sampling and resource allocation.

Low-altitude networks, while dynamic, are often predictable \cite{JunBowHaoShu:A25}. For instance, \acpl{uav} performing aerial inspection or delivery typically follow trajectories predetermined by their mission requirements, making their future channel conditions predictable. Some preliminary results have shown that this predictive information can enable proactive strategies to improve transmission efficiency, provided that a given data generation process is assumed \cite{LiChe:J24a, LiChe:J24b}. However, when data generation is optimized, the sampling timing and transmission strategy are tightly coupled, whereas exhaustive search for the best sampling timing is exponential in complexity.

In addition, the low-altitude and terrestrial networks are tightly coupled through dominant \ac{los} channels, creating a direct conflict between their service objectives. Consequently, guaranteeing data freshness necessitates a trade-off between two distinct operational costs: the \ac{uav} can either expend more of its own energy to increase transmission power, impacting its operational endurance, or it can occupy more of the shared channel resources to increase the spectrum for transmission, reducing its availability for terrestrial users. Therefore, a critical challenge is to find an optimal balance between these two costs, selecting a strategy that satisfies the timeliness requirement in the most efficient way for the coexistence of the two networks.

In this work, we formulate a bi-objective optimization problem to characterize the fundamental trade-off between aerial energy consumption and \ac{rb} usage, under a strict constraint on aerial timeliness. The goal is to find the complete Pareto frontier of all optimal sampling and communication strategies. However, characterizing the complete Pareto frontier is a non-trivial task, as classical methods, such as weighted-sum approaches and heuristic algorithms, cannot guarantee finding all optimal trade-off points \cite{JakBlu:J14,ChoWanCheCha:J17}. This difficulty is compounded by the inherent complexity of the underlying joint optimization problem itself, which is non-convex, features an unknown number of variables (dimension-unknown), and involves mixed-integer constraints.

To tackle these challenges, we propose a predictive, status-aware framework with the following key contributions
\begin{itemize}
    \item We propose a two-layer optimization framework, proven to find all Pareto optima, that first transforms the bi-objective problem into a single-objective equivalent and then decomposes it into an inner problem for communication strategy design and an outer problem for sampling control.
    \item A graph-based control algorithm is proposed to find the Pareto optimum with low complexity, linear in the horizon and near-quadratic in the \ac{rb} budget. 
\end{itemize}

\section{System Model\label{sec:System-Model}}
\begin{figure}
\begin{centering}
\includegraphics[width=1\columnwidth]{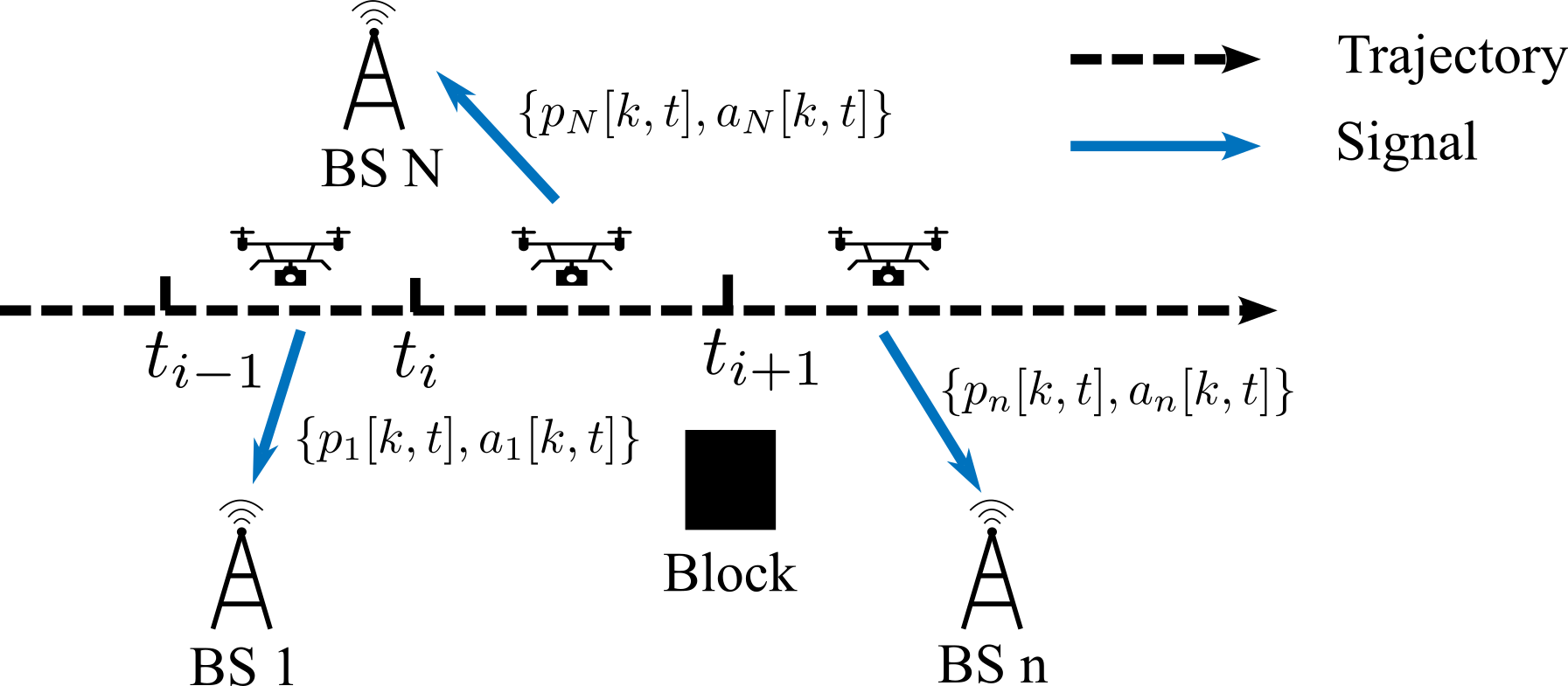}
\par\end{centering}
\caption{\ac{uav} telemetry system model. The \ac{uav} symbols along the trajectory illustrate the \ac{uav}'s positions at different time instants. The \ac{uav} reports its state information or sensing data to a fusion center through \acpl{bs}.}
\label{fig:system_model}
\end{figure}

Consider a \ac{uav} telemetry system, as depicted in Figure~\ref{fig:system_model}. The system consists of $1$ \ac{uav} indexed by $0$, and $N$ \acpl{bs} indexed by $\mathcal{N}=\{1,\cdots, N\}$. The \ac{uav} is tasked with transmitting on-board sensory data to a fusion center via the terrestrial network ({\em i.e.}, the BSs). Timely delivery of this information is crucial, as the fusion center relies on the most recent and relevant data for reliable analysis and informed decision-making. Meanwhile, the terrestrial network shall also maintain stable service for the ground users.

To achieve fair coexistence of aerial and terrestrial traffic, we aim to design a resource management strategy that jointly determines (i) the optimal timing for sampling and transmission of aerial data, and (ii) the optimal power and spectrum allocation decisions. We leverage predictive channel models so that these decisions are optimal over a long-term planning horizon.

\subsection{Predictive Channel Model}
We consider a slotted \ac{ofdm} system, where time is divided into slots indexed by $t \in \mathcal{T}=\{1,\cdots,T\}$, and the available spectrum is partitioned into $K$ orthogonal \acpl{rb}, indexed by $k \in \mathcal{K}=\{1,\cdots,K\}$. Generally, the wireless channel from the \ac{uav} to \ac{bs} $n$ on \ac{rb} $k$ at time $t$ can be modeled as
\begin{equation}
h_{n}\left[k,t\right]=g_{n}\left[k,t\right]\xi_{n}\left[k,t\right],
\label{eq:channel_model}
\end{equation}
where $g_{n}[k,t]$ is the large-scale channel gain (e.g., path loss and shadowing), and $\xi_{n}[k,t]\sim\text{Gamma}(\kappa_{n}[k,t],1/\kappa_{n}[k,t])$ captures small-scale fading with unit mean \cite{LiChe:J24b,Nak:B60}.

The predictive channel model is built on two key enablers.
\begin{enumerate}
    \item advanced channel sensing techniques, such as radio maps and digital twins \cite{XinChe:J24,Sunche:J24,WanZhaNieYu:M25}, which provide a 3D representation of the wireless propagation environment and offer spatially resolved channel statistics between the UAV and ground BSs; and 
    \item high-precision \ac{uav} control, which allows the UAV to follow pre-determined trajectories $\{(t, \mathbf{p}_{0}[t])\}_{t\in\mathcal{T}}$ with minimal deviation \cite{JunBowHaoShu:A25}.
\end{enumerate}
Consequently, the UAV’s motion traces a one-dimensional slice through the 3D channel field, yielding a time-indexed channel profile that can be predicted in advance. Let
\begin{equation}
\left[g_{n}\left[k,t\right],\kappa_{n}\left[k,t\right]\right]=\Xi\left(\mathbf{p}_{0}\left[t\right],\mathbf{p}_{n}\left[t\right]\right),\,\, t\in\mathcal{T}
\end{equation}
denote the radio map between the UAV and BS $n$ along the trajectory. The resulting predictive channel model can then be expressed as
\begin{equation}
h_{n}\left[k,t\right]\sim\text{Gamma}\left(\kappa_{n}\left[k,t\right],g_{n}\left[k,t\right]/\kappa_{n}\left[k,t\right]\right).
\end{equation}

\subsection{Transmission Model}

Denote the allocation of \ac{rb} $k$ at the time slot $t$ for the communication from node $0$ to \ac{bs} $n$ as $a_{n}[k,t]\in\{0,1\}$. For each $(k,t)$ \ac{rb}, the node $0$ is allowed to transmit to at most one \ac{bs}, that is
\begin{equation}
\sum_{n\in\mathcal{N}}a_{n}\left[k,t\right]\le1,\,\,\forall k\in\mathcal{K},t\in\mathcal{T}.\label{eq:rb_c}
\end{equation} 
Let $p_{n}\left[k,t\right]$ denote the transmit power. The total transmission power is limited to the threshold $\bar{p}$, leading to the sum-power constraint
\begin{equation} \label{eq:p_def}
\sum_{k\in\mathcal{K}}\sum_{n\in\mathcal{N}}p_{n}\left[k,t\right]\le\bar{p},\,\,\forall t\in\mathcal{T}.
\end{equation}

We now derive the data throughput. The \ac{snr} for the link from node $0$ to \ac{bs} $n$ on \ac{rb} $k$ at time $t$ is given by $\gamma_{n}\left[k,t\right]=p_{n}\left[k,t\right]h_{n}\left[k,t\right]/\delta^{2}$, where $\delta^{2}$ is the noise power. Assuming perfect Doppler compensation through advanced techniques\cite{LuZen:J24}, then, the channel capacity from node $0$ to node $n$ at time $t$ for block $k$ is modeled as
\begin{equation}
c_{n}\left[k,t\right]=B\log_{2}\left(1+\gamma_{n}\left[k,t\right]\right)\label{eq:def_c}
\end{equation}
where $B$ is the single-block bandwidth. Finally, the total data
throughput over a time interval $(t, t^\prime)$ aggregated across all receiving \acpl{bs} and allocated \acpl{rb} is given by
\begin{equation}
\upsilon\left(t,t^\prime\right)=\sum_{n\in\mathcal{N}}\sum_{k\in\mathcal{K}}\sum_{\tilde{t}=t}^{t^\prime-1}c_{n}\left[k,\tilde{t}\right]a_{n}\left[k,\tilde{t}\right].\label{eq:thp_def}
\end{equation}

\subsection{Metrics for Coexistence}

\subsubsection{Timeliness requirement for aerial traffic}
In this work, we use the \ac{aoi} metric to quantify the freshness of information received from the \ac{uav}. Let $s[t]\in\{0,1\}$ denote the update-success indicator for node $0$ at the end of the slot $t$, and $t_{0}$ denote its last sampling time. The \ac{aoi} at the fusion center is recursively defined as
\begin{equation}
\tau\left[t+1\right] \triangleq \begin{cases}
t-t_{0}, & s[t]=1,\\
\tau\left[t\right]+1, & s[t]=0.
\end{cases}
\end{equation}
A transmission attempt is successful if the expected delivered payload accumulated since the previous success meets a quality threshold $\bar{\upsilon}$,
{\em i.e.}, 
\begin{equation}
s\left[t\right]=\mathbb{I}\left\{ \mathbb{E}\left\{ \upsilon\left(t_{0},t\right)\right\} \ge\bar{\upsilon}\right\} .\label{eq:aoi_I}
\end{equation}
Here, the expectation is with respect to the channel $h_n[k,t]$ for all $n\in\mathscr{N}$, $k\in\mathcal{K}$ and $t\in[t_0,\cdots, t-1]$.
For timeliness, we impose a hard constraint on the peak information age, {\em i.e.},
\begin{equation}
\tau\left[t\right]\le\bar{\tau},\,\,\forall t\in\mathcal{T}.\label{eq:aoi_def}
\end{equation}

\subsubsection{Fairness requirement for coexistence}
For fairness between aerial and terrestrial services, we aim to regulate the aerial load so that no \ac{bs} experiences excessive spectrum occupation by \ac{uav} traffic over time.
To this end, we characterize the temporal load level at \ac{bs} $n$ by its worst-case load occupied by aerial traffic, i.e., $l_{n}\triangleq\max_{t\in\mathcal{T}}\sum_{k\in\mathcal{K}}a_{n}\left[k,t\right]$. To further promote spatial fairness across the network, we define the spatiotemporal load cap as
\begin{equation}
\theta\triangleq\max_{n\in\mathcal{N}}l_{n} = \max_{n\in\mathcal{N},t\in\mathcal{T}}\sum_{k\in\mathcal{K}}a_{n}\left[k,t\right].\label{eq:def_stable}
\end{equation}

\subsubsection{Energy efficiency for aerial traffic}
In contrast to ground \acpl{bs}, which have a stable and sufficient energy supply, the energy consumption at the \ac{uav} is tightly constrained by its limited onboard battery capacity, making energy efficiency a critical design consideration. Formally, we define the energy consumption as
\begin{equation}
E\triangleq\sum_{n\in\mathcal{N},k\in\mathcal{K},t\in\mathcal{T}}a_{n}\left[k,t\right]p_{n}\left[k,t\right].\label{eq:def_ee}
\end{equation}


\subsection{Age-Aware Sampling Controller}
 The \ac{aoi} serves not merely as a timeliness constraint; it also allows direct control of information generation \cite{luo2025information}. Given the predictive communication model over the entire planning horizon, we aim to control the sampling timing of sensory data so as to achieve our goal with minimal payload.
 
Mathematically, denote $t_{i}$ as the sampling instant (\emph{i.e.}, data generation time) of the $i$th status packet of the node $0$. The sequence of sampling instants is $\mathbf{t}=\{t_{1},t_{2},\cdots,t_{I}\}$, where
\begin{equation}
    1\leq t_{i} \le T,\,\,t_{i}\in\mathcal{T}, i\in\mathcal{I} = \{1,\cdots,I\}, \label{eq:aoi_c_s3}
\end{equation}
and $I$ is the total number of sampling events. Without loss of generality, we assume an initial sample is taken at the beginning of the first slot $t_0=1$.

Due to the \ac{aoi} constraint (\ref{eq:aoi_def}), the sampling interval cannot be larger than $\bar{\tau}$, that is,
\begin{equation}
1\le t_{i+1}-t_{i}\le\bar{\tau},\,\,\forall i\in\mathcal{I}.\label{eq:aoi_c_s1}
\end{equation}
We set $t_{I+1}=T+1$ to represent the end time of transmission for the $I$th status update. Recall that the transmission is successful if the link throughput satisfies~\eqref{eq:aoi_I}. Accordingly, we impose the following expected throughput constraint between any two consecutive sampling instances
\begin{equation}
\mathbb{E}\left\{ \upsilon\left(t_{i},t_{i+1}\right)\right\} \ge\bar{\upsilon},\,\,\forall i\in\mathcal{I}.\label{eq:aoi_c_s2}
\end{equation}
As a result, the \ac{aoi} constraint (\ref{eq:aoi_def}) is equivalent to \eqref{eq:aoi_c_s3}-\eqref{eq:aoi_c_s2}.

\section{Problem Formulation}
The goal is to maintain the timeliness of aerial data while satisfying coexistence requirements. To this end, we jointly optimize the sampling timing sequence $\mathbf{t}=\{t_{1},t_{2},\cdots,t_{I}\}$, transmit power $\mathbf{P}=\{p_{n}\left[k,t\right]\}_{n\in\mathcal{N},k\in\mathcal{K},t\in\mathcal{T}}$, and \ac{rb} allocation $\mathbf{A}=\{a_{n}\left[k,t\right]\}_{n\in\mathcal{N},k\in\mathcal{K},t\in\mathcal{T}}$, to minimize the spatiotemporal load cap $\theta$ and the energy consumption $E$ in the \emph{Pareto} sense under hard \ac{aoi} constraint. The bi-objective optimization problem is formulated as follows
\begin{align}
\mathscr{P}1:\underset{\mathbf{t},\mathbf{P},\mathbf{A}}{\text{minimize}} & \ \left\{ \theta,E\right\}  \nonumber \\
\text{subject to} & \ \mathbb{E}\left\{ \upsilon\left(t_{i},t_{i+1}\right)\right\} \ge\bar{\upsilon},\forall i \label{eq:thp_P1_c1}\\
& \ \mathbf{t}\in \Upsilon, \mathbf{P} \in \mathcal{P}(\mathcal{T}),\mathbf{A}\in \mathcal{A}(\mathcal{T})\label{eq:basic_P1_c1}.
\end{align}
Here, $\Upsilon$, $\mathcal{P}$, and $\mathcal{A}$ denote the feasible spaces of sampling time, power allocation, and \ac{rb} scheduling over the entire planning horizon $\mathcal{T}$, respectively. Specifically, from (\ref{eq:aoi_c_s3}) and (\ref{eq:aoi_c_s2}), the space of sampling time is
\[\Upsilon = \left\{\left\{t_i\in\mathcal{T}\right\}:1\le t_{i+1}-t_{i}\le\bar{\tau},\forall i\in \mathcal{I}(I), I\in\mathbb{R}^+\right\}.\]
From (\ref{eq:p_def}), the space of the power allocations is 
\[ \mathcal{P}(\mathcal{T})=\big\{\{p_n[k,t]\ge 0\} :\sum_{k\in\mathcal{K}}\sum_{n\in\mathcal{N}}p_{n}\left[k,t\right]\le\bar{p},\forall t\in\mathcal{T}\big\}, \]
and from (\ref{eq:rb_c}), the \ac{rb} allocation space $\mathcal{A}(\mathcal{T})$ is 
\begin{align}
\big\{\{a_n[k,t]\in\left\{0,1\right\}\} :\sum_{n\in\mathcal{N}}a_{n}\left[k,t\right]\le1,\forall k\in\mathcal{K},t\in\mathcal{T}\big\}.\nonumber
\end{align}

The solution to $\mathscr{P}1$ in the Pareto sense is the determination of the complete Pareto frontier that consists of all Pareto-optimal strategies (see Definition~\ref{def:patero} below). That is, there is no single `best' strategy in Pareto optimization; instead, each point represents a trade-off between the two competing objectives.

\begin{defn}\label{def:patero}
(Pareto optimality.) A strategy $\pi = (\mathbf{t}, \mathcal{P}, \mathcal{A})$ is Pareto-optimal if there is no feasible strategy $\pi^\prime$ such that $\theta(\pi^\prime) \leq \theta(\pi)$ and $E(\pi^\prime) \leq E(\pi)$ with at least one strict inequality. The Pareto frontier is the set of all Pareto-optimal points in the objective space.
\end{defn}
\textit{Remark:} Problem $\mathscr{P}1$ presents several fundamental challenges:
(i) determining the complete Pareto frontier is generally non-trivial, as illustrated in Figure~\ref{fig:Pareto_illu};
(ii) the underlying optimization is a non-convex mixed-integer problem, which is computationally demanding; and
(iii) optimizing the variable $\mathbf{t}$ is particularly difficult since its dimension $I$ is not fixed, leading to numerous isolated local optima. Mathematically, varying $I$ corresponds to a mode-switching problem.

In the following sections, we shall derive some convenient properties of the Pareto frontier, building on which we develop an efficient graph-based method to solve the problem.


\section{Pareto Analysis and Problem Decomposition\label{sec:Problem-Reformulation}}
This section characterizes the complete Pareto frontier. We first show that, for any given load cap $\theta$, $\mathscr{P}1$ reduces to a single-objective problem $\mathscr{P}2$ that seeks to minimize the energy consumption $E^*(\theta)$ for a given $\theta$. A strong result in Proposition~\ref{prop:pareto_front} shows that $(\theta, E^*(\theta))$ in a specific domain characterizes the complete Pareto frontier. This result holds for any strictly increasing functions of the objectives $\theta$ and $E$. Moreover, $\mathscr{P}2$ can be decomposed into a two-layer problem that can be solved efficiently.

\subsection{Pareto Analysis}
Consider $\theta$ as an optimization variable, then the spatiotemporal load cap given by \eqref{eq:def_stable} can be equivalently represented by the following epigraph constraint~\cite{Boy:B04}
\begin{equation}
\sum_{k\in\mathcal{K}}a_{n}\left[k,t\right]\le \theta,\,\forall n\in\mathcal{N},t\in\mathcal{T}.\label{eq:theta_c} 
\end{equation}
Then, Problem $\mathscr{P}\text{1}$ reduces to a single-objective problem 
\begin{equation}
\mathscr{P}2: \min_{\left\{ \mathbf{t},\boldsymbol{\mathcal{P}},\boldsymbol{\mathcal{A}}\right\} \in\mathcal{F}\left(\theta\right)} \sum_{n\in\mathcal{N},k\in\mathcal{K},t\in\mathcal{T}}a_{n}\left[k,t\right]p_{n}\left[k,t\right],\notag
\end{equation}
where $\mathcal{F}\left(\theta\right)$ denotes the set of feasible strategies for a fixed 
load cap $\theta$, where
\begin{equation}
\mathcal{F}\left(\theta\right)\triangleq\left\{ \left\{ \mathbf{t},\boldsymbol{\mathcal{P}},\boldsymbol{\mathcal{A}}\right\} \text{: \eqref{eq:thp_P1_c1}-\eqref{eq:theta_c}}\right\} .\label{eq:def_F_theta}
\end{equation}

Let $E^*(\theta)$ denote the optimal value of $\mathscr{P}2$. The following Proposition establishes the complete Pareto frontier of $\mathscr{P}1$.

\begin{figure}
\begin{centering}
\includegraphics[width=\columnwidth]{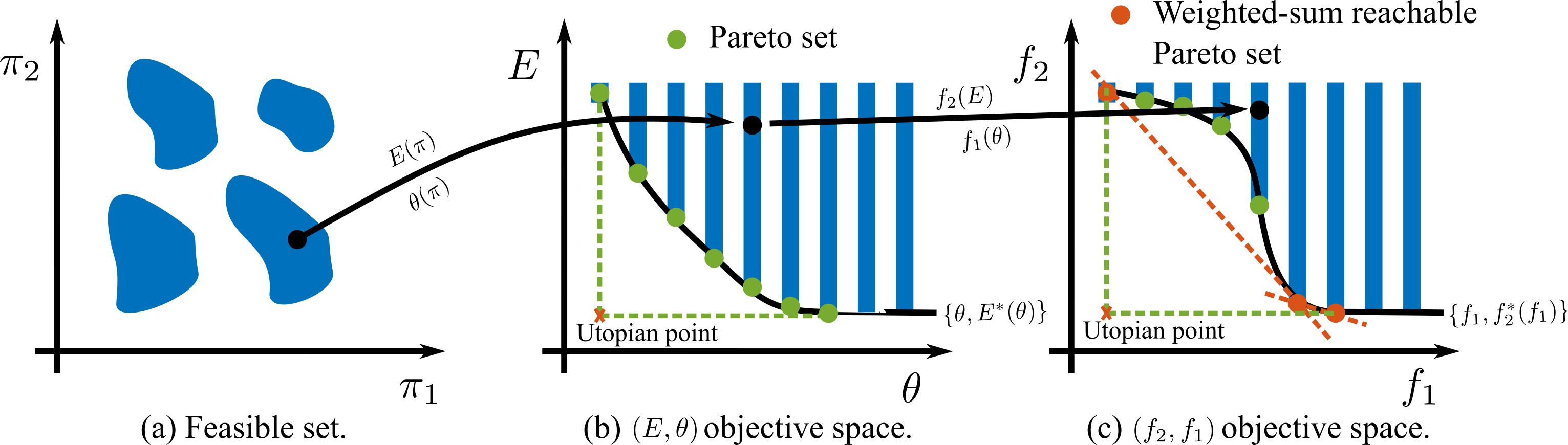}
\par\end{centering}
\caption{\label{fig:Pareto_illu}Illustration of Pareto optimality for a two-variable, two-objective optimization problem. (a) The nonconvex feasible set. (b) Bi-objective $(E,\theta)$ space, where the feasible space and the objective space are non-convex and non-continuous, thereby classical heuristic algorithms are challenging to find the complete Pareto frontier. (c) Bi-objective $(f_2,f_1)$ space, where the Pareto frontier is non-concave, thereby, the weighted-sum method cannot guarantee the discovery of all Pareto optima.}
\end{figure}

\begin{prop}
\label{prop:pareto_front}(Pareto frontier.) The set  
\begin{align}
    \mathcal{C}\triangleq\{(\theta,E^*(\theta)):\theta\in\{\underline{\theta},\cdots\overline{\theta}\}\}
\end{align}
is the Pareto frontier of $\mathscr{P}1$, where $\underline{\theta}\triangleq\theta^{*}$, 
\begin{equation}
\overline{\theta}\triangleq\min\{\theta\in\mathbb{Z}^{+}:E^*(\theta)=E^{*}\},\label{eq:def_theta_up}
\end{equation}
and $\{\theta^{*},E^{*}\}$ is the utopian point of the frontier (see Figure~\ref{fig:Pareto_illu}).
\end{prop}


This result holds for any strictly increasing scalarization of the bi-objective problem, which we formalize below.
\begin{cor}
\label{cor:pareto_front}For any $f_{1}(\cdot)$ and $f_{2}(\cdot)$, with $f_{1}^{\prime}(\cdot)>0$ and $f_{2}^{\prime}(\cdot)>0$. The set $\mathcal{C}_{f}\triangleq\{(f_{1}(\theta),f_{2}(E^*(\theta))):\theta\in\{\underline{\theta},\cdots\overline{\theta}\}\}$ is the Pareto frontier of the problem with the objective $\{f_{1}(\theta),f_{2}(E)\}$
under the same constraints as Problem $\mathscr{P}1$.
\end{cor}

\subsection{Problem Decomposition}

It is observed from problem $\mathscr{P}2$ that the variables are coupled over time $t$ only by the objective function and constraint (\ref{eq:thp_P1_c1}). Therefore, given any feasible sampling variable $\mathbf{t}$, problem $\mathscr{P}2$ can be decomposed into $I$ parallel resource allocation problems. Denote $\mathcal{T}_{i}\triangleq\{t_{i},t_{i}+1,\cdots,t_{i+1}-1\}$ as the transmission interval of the $i$th sampling data, we can convert problem $\mathscr{P}2$ to a two-layer problem.
\begin{prop}
\label{prop:opt_trans}(Decomposition of $\mathscr{P}2$.) Problem $\mathscr{P}2$ is equivalently transformed into the following outer
subproblem
\[
\mathscr{P}\text{2-1}:\quad\underset{\mathbf{t}}{\text{min}}\ \sum_{i\in\mathcal{I}}E^{*}\left(t_{i},t_{i+1}\right)\text{ s.t. } \mathbf{t}\in \Upsilon,
\]
where $E^{*}(t_{i},t_{i+1})$ is the solution to the inner subproblem 
\begin{align}
\mathscr{P}\text{2-2}: \underset{\mathbf{\pi}_{i}}{\text{min}} & \ \sum_{n\in\mathcal{N},k\in\mathcal{K},t=t_{i}}^{t_{i+1}-1}a_{n}\left[k,t\right]p_{n}\left[k,t\right] \nonumber\\
\text{s.t.} & \ \text{(\ref{eq:thp_P1_c1}) for } i, \sum_{k\in\mathcal{K}}a_{n}\left[k,t\right]\le \theta,\,\forall n\in\mathcal{N},t\in\mathcal{T}_i \nonumber\\
& \  \left\{p_m[k,t]\right\}\in \mathcal{P}(\mathcal{T}_{i}), \left\{a_m[k,t]\right\}\in \mathcal{A}(\mathcal{T}_{i})\nonumber
\end{align}
where $\mathbf{\pi}_{i}=\{p_{n}\left[k,t\right],a_{n}\left[k,t\right]\}_{n\in\mathcal{N},k\in\mathcal{K},t\in\mathcal{T}_{i}}$.
\end{prop}

\section{Graph-Based Algorithm for Optimal Control}
This section proposes a framework to solve Problem $\mathscr{P}\text{2-1}$ and Problem $\mathscr{P}\text{2-2}$. First, $\mathscr{P}\text{2-2}$ can be transformed into a convex problem by relaxing $a_{n}\left[k,t\right]\in[0,1]$ and introducing an auxiliary variable $\phi_{n}[k,t]=a_{n}\left[k,t\right]p_{n}\left[k,t\right]$. Consequently, $\mathscr{P}\text{2-2}$ can be solved with an optimality guarantee in $\mathcal{O}((t_{i+1}-t_{i})\theta\log_{2}((t_{i+1}-t_{i})\theta))$ complexity \cite{LiChe:J24b}. Thus, the remaining task is to solve the integer programming
problem $\mathscr{P}\text{2-1}$. The proofs can be found in \cite{LiLuoThePap:A25}.

\subsection{Graph-Based Outer Solution} \label{subsec:graph}

\begin{figure}
\begin{centering}
\includegraphics[width=\columnwidth]{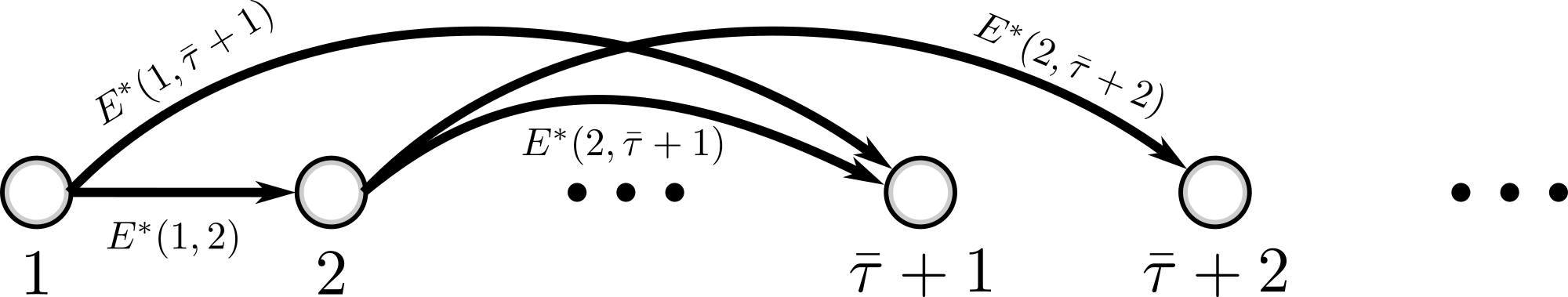}
\par\end{centering}
\caption{\label{fig:Sample-timing-graph}Illustration of the timing-control graph, where each vertex represents a possible sampling instant, each directed edge denotes a transmission during the interval between the two sampling instants, and the edge weight indicates the optimal energy consumption for the transmission induced by the edge.}
\end{figure}

We construct a timing-control graph $\mathscr{G}=\{\mathbf{v},\mathbf{e},\mathbf{w}\}$, as shown in Figure~\ref{fig:Sample-timing-graph}. Here, the vertex set $\mathbf{v}=\{1,\cdots,T+1\}$ represents all admissible sampling timings; the terminal node $T+1$ marks the end boundary of the horizon so that each transmission interval is $\{t_{i},t_{i}+1,\cdots,t_{i+1}-1\}$.
The direct edge set $\mathbf{e}=\{(v_{i},v_{j})\}$ represents the transmission event, where $v_{i},v_{j}\in\mathbf{v}$ for satisfying constraint $t_i\in\mathcal{T}$ and $1\le t_{i}-t_{j}\le\bar{\tau}$ for satisfying \ac{aoi} constraint. The weight set $\mathbf{w}=\{w_{i,j}\}$ represents the transmission cost, defining as the optimal solution to $\mathscr{P}\text{1-2}$ with sampling interval $t_{i}$ and $t_{j}$, that is $w_{i,j}=E^{*}(t_{i},t_{j})$.

With this construction, a feasible sampling sequence $1=t_{0}<t_{1}<\cdots<T_{I+1}=T+1$ corresponds to a directed path 
\[
1\to t_{1}\to\cdots\to T_{I+1}=T+1
\]
whose total cost equals $\sum_{i=1}^{I}E^{*}(t_{i},t_{i+1})$.
\begin{prop}
\label{prop:opt_P21}(Equivalence of $\mathscr{P}\text{2-1}.$) The optimal solution to $\mathscr{P}\text{2-1}$ is the shortest path from node $1$ to node $T+1$ in $\mathscr{G}$.
\end{prop}

Consequently, problem $\mathscr{P}\text{2-1}$ can be solved using a classical shortest path algorithm on a weighted, directed graph.

\subsection{Graph-Based Algorithm}

The algorithm is summarized in Algorithm \ref{alg:g_control_alg}, which proceeds in two phases: construct the timing-control graph $\mathscr{G}$ with edge weights given by the optimal interval energy $E^{*}(\cdot,\cdot)$; find the shortest path from node $1$ to node $T+1$.

\begin{algorithm}
\# Input: $\theta$, $g_n[k,t]$ and $\kappa_n[k,t]$
\begin{enumerate}
\item Construct the graph $\mathscr{G}$ based on Figure~\ref{fig:Sample-timing-graph}
and the weights are calculated by solving $\mathscr{P}\text{2-2}$
according to \cite{LiChe:J24b}.
\item Shortest path algorithm from node $1$ to node $T+1$ to find the
optimal $\mathbf{t}^{*}$. 
\item Calculate $\mathbf{P}_{n}^{*}$, $\mathbf{A}_{n}^{*}$, $E^{*}(\theta)$, according to $\mathbf{t}^{*}$.
\end{enumerate}
\# Output: $\mathbf{P}_{n}^{*}$, $\mathbf{A}_{n}^{*}$, $\mathbf{t}^{*}$,
$E^{*}(\theta)$

\caption{\label{alg:g_control_alg}Graph-based control algorithm}
\end{algorithm}

\subsubsection{Optimality Analysis}
First, the subproblem $\mathscr{P}\text{2-2}$ can be solved with an optimality guarantee as established in \cite{LiChe:J24b}. Therefore, each edge weight $w_{i,j}$ in the constructed graph $\mathscr{G}$ represents the optimal interval energy $E^{*}(t_{i},t_{j})$. Next, according to Proposition \ref{prop:opt_P21}, the shortest path in $\mathscr{G}$ yields the optimal solution to the outer subproblem $\mathscr{P}\text{2-1}$. Finally, by Proposition \ref{prop:opt_trans},
$\mathscr{P}\text{2-1}$ and $\mathscr{P}\text{2-2}$ together are equivalent to the original problem $\mathscr{P}2$. Accordingly, the
output of Algorithm \ref{alg:g_control_alg} achieves the globally optimal solution to $\mathscr{P}2$.

\subsubsection{Complexity Analysis}
The complexity for all Pareto frontier searching is $\mathcal{O}(T\theta^2\bar{\tau}^{2}\log(\bar{\tau}\theta))$, where for each $\theta$, the complexity for the shortest path over a directed graph with nonnegative weights is $\mathcal{O}(T\bar{\tau}+T)=\mathcal{O}(T\bar{\tau})$, and the complexity for calculating all the weights $E^{*}(t_{i},t_{j})$ in $\mathscr{G}$ is $\mathcal{O}(T\theta\bar{\tau}^{2}\log\bar{\tau}\theta)$.
\begin{figure}[ht]
\begin{centering}
\includegraphics[width=0.8\columnwidth]{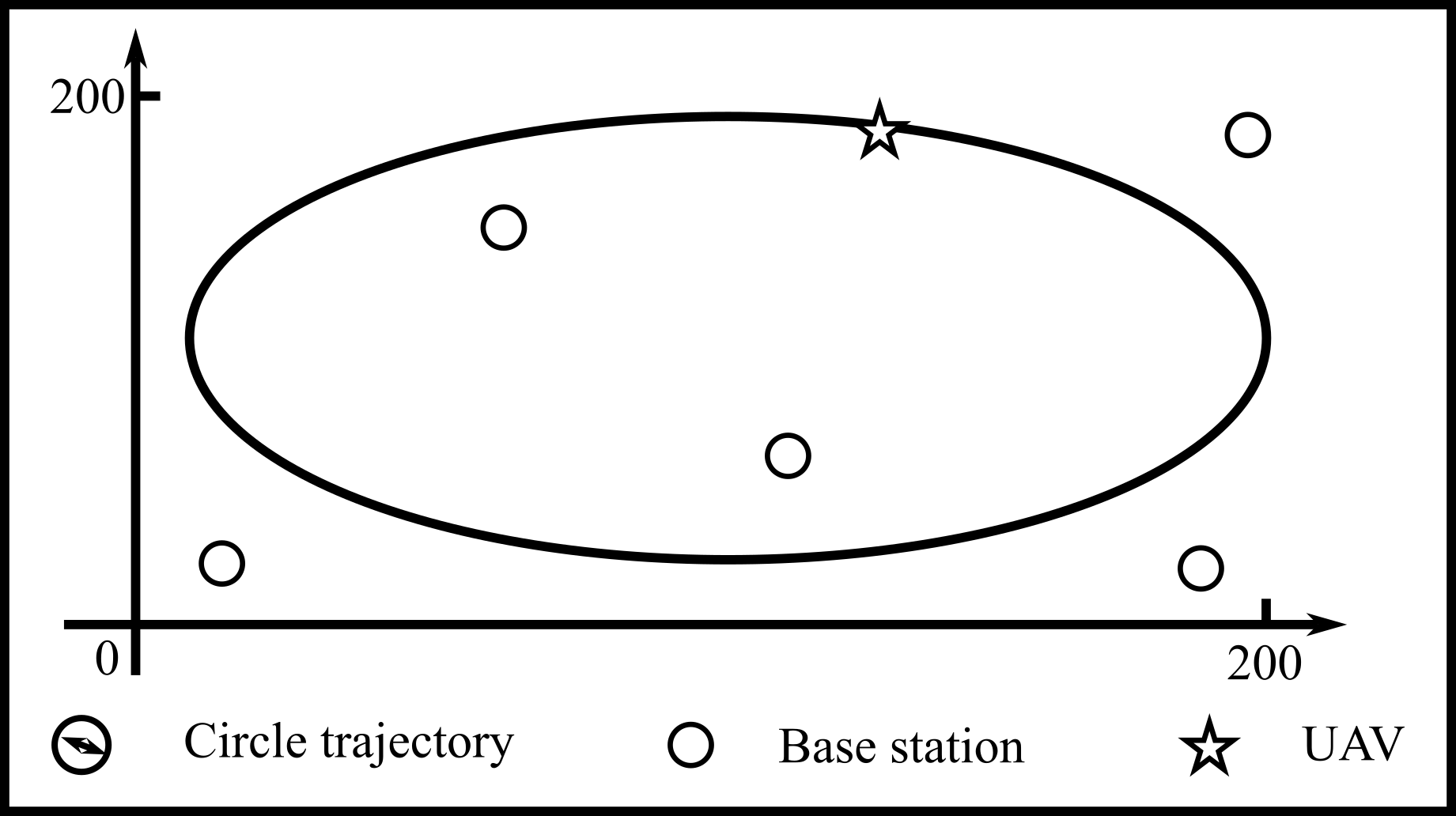}
\par\end{centering}
\caption{\label{fig:Simulation_layout_illustration}The simulation layout with $N=5$ \acpl{bs} and one patrol \ac{uav}.}
\end{figure}

\section{Simulation Results\label{sec:Simulation}}
We consider a patrol system in which a \ac{uav} follows a circular trajectory to monitor a $200\times200,\text{m}^2$ area, as illustrated in Figure~\ref{fig:Simulation_layout_illustration}. The \ac{uav} operates at an altitude of $50$ m with a flight speed of 6 m/s. On the ground, $N$ \acpl{bs} are deployed, with their positions randomly generated. The channels are realized according to (\ref{eq:channel_model}). Similar to \cite{LiChe:J24b}, the shape parameters $\kappa$ of the Gamma distribution of small-scale fading $\xi$ are set randomly in [1, 30], and the large-scale fading $g$ includes path loss and shadowing, where the path loss is generated by 3GPP \ac{umi} model \cite{TR36814} and the channel block state is generated by \ac{los} probability model \cite{MozSadBen:J17}. In contrast, the shadowing is modeled by a log-normal distribution.

\begin{figure}[t!]
\begin{centering}
\includegraphics[width=1\columnwidth]{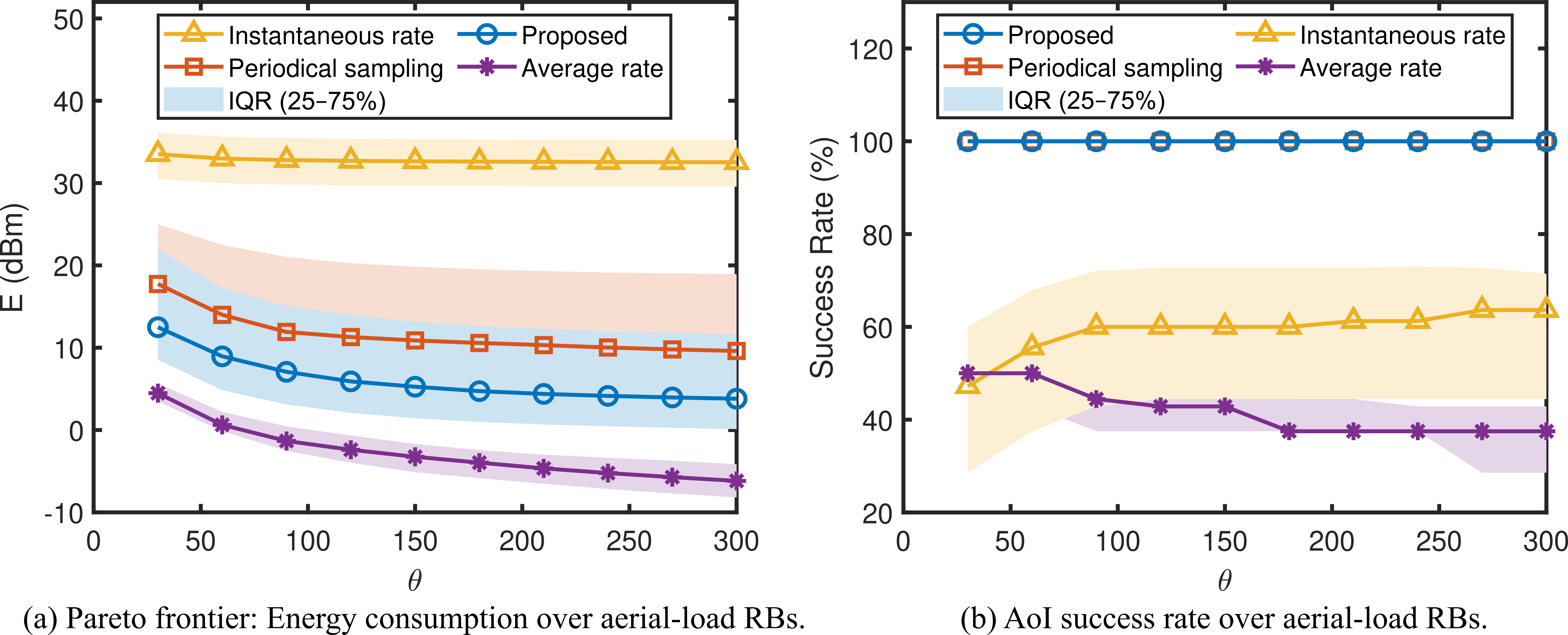}
\par\end{centering}
\caption{\label{fig:pareto_opt}Energy consumption and \ac{aoi} success rate over the aerial-load \acpl{rb}, where the solid lines with markers denote the median values, while the shaded areas indicate the interquartile range (IQR, 25th-75th percentile).}
\end{figure}

We compare our performance with the following three baselines (two with no sampling control and one with sampling control but no sampling optimization). \emph{1) Instantaneous rate \cite{MatShe:J25}:} Trade-off spatiotemporal load cap and energy efficiency under the piecewise rate constraint, that is, $c_{m}(t)\ge S/\bar{\tau}$, $\forall t$. \emph{2) Average rate \cite{ZhaLiRonZen:J25}:} Trade-off spatiotemporal load cap and energy efficiency under the average rate constraint, that is, $\sum_{t}c_{m}(t)/T\ge S/\bar{\tau}$. \emph{3) Periodical sampling:} The sampling time is fixed as $t_{k}=(k-1)\bar{\tau}$, while the resource allocation strategy follows the proposed schemes.

Figure~\ref{fig:pareto_opt} (a) illustrates the Pareto frontiers of all compared schemes. It can be observed that the non-predictive scheme (instantaneous rate) is significantly suboptimal, by more than 20 dB, compared with the other three predictive schemes, highlighting the importance of predictive information in planning. Moreover,
although both the periodical and proposed status-aware schemes exploit predictive information, the latter achieves substantial performance improvement by adaptively controlling the sampling instants. For example, given an energy budget of $E=10\text{ dBm}$, the periodical sampling method requires approximately 300 \acpl{rb} to complete the task. In contrast, the proposed status-aware method needs only 50 \acpl{rb}.

Figure~\ref{fig:pareto_opt} (b) illustrates the fulfillment of the aerial timeliness requirement, showing that the achieved \ac{aoi} remains below the maximum tolerable threshold. Firstly, although the average-rate method exhibits superior energy efficiency and \ac{rb} utilization compared with other baselines, it fails to satisfy the \ac{aoi} requirement consistently. This is because the average-rate predictive scheme optimizes the long-term
average throughput rather than per-interval performance; thus, intervals coinciding with poor channel conditions remain underserved, leading to unstable \ac{aoi} satisfaction even when $\theta$ increases. Similarly, the instantaneous rate method, despite consuming the most resources, also fails to guarantee the \ac{aoi} constraint. Without prediction, it cannot anticipate deep fades; once a poor channel interval occurs, the per-interval requirement cannot be met. In contrast, the proposed predictive status-aware controller maintains 100\% \ac{aoi} satisfaction across all cases.


\section{Conclusion\label{sec:Conclusion}}

In this paper, we designed a framework that jointly controls the data generation process (via sampling timing) and data transmission (via resource allocation). This approach successfully guarantees strict timeliness requirements while simultaneously minimizing both the \ac{uav}'s energy consumption and its impact on the terrestrial network. The cornerstone of our approach is a novel two-layer graph-based algorithm, which we proved can efficiently characterize the complete Pareto frontier with a complexity of $\mathcal{O}(T\theta^2\bar{\tau}^{2}\log(\bar{\tau}\theta))$. The efficacy of this framework is demonstrated by significant resource savings, achieving up to a six-fold reduction in RB utilization and a 6 dB reduction in energy consumption in our simulations.

\bibliographystyle{IEEEtran}
\bibliography{IEEEabrv,StringDefinitions,BL}

\clearpage

\appendices

\section{Proof of Proposition \ref{prop:pareto_front}\label{sec:proof_prop_pareto_front}}

We will prove that any point in $\mathcal{C}$ is Pareto-optimal and all Pareto-optimal solutions lie on $\mathcal{C}$. To establish this result, we first prove two key lemmas concerning the monotonicity of the $E^*(\theta)$.

\subsection{Monotonicity of $E^*(\theta)$}

\begin{lem}
\label{lem:monotonic_E_theta}
For any $\theta_{1}<\theta_{2}$, $E^*\left(\theta_{1}\right)\ge E^*\left(\theta_{2}\right)$.
\end{lem}
\begin{IEEEproof}
Given any $\theta_{1}<\theta_{2}$. If $\mathcal{F}\left(\theta_{1}\right)=\emptyset$, we have $\mathcal{F}\left(\theta_{1}\right)\subseteq\mathcal{F}\left(\theta_{2}\right)$. Otherwise, for any $\mathcal{F}\left(\theta_{1}\right)\neq\emptyset$,
for any $\{\mathbf{t}^{\prime},\boldsymbol{\mathcal{P}}^{\prime},\boldsymbol{\mathcal{A}}^{\prime}\}\in\mathcal{F}\left(\theta_{1}\right)$,
the load cap constraint \eqref{eq:theta_c} ensures $\sum_{k\in\mathcal{K}}a_{n}^{\prime}\left[k,t\right]\le\theta_{1},\forall n,t.$
Since $\theta_{1}<\theta_{2}$, it follows that $\sum_{k\in\mathcal{K}}a_{n}^{\prime}\left[k,t\right]\le\theta_{2},\forall n,t.$
Moreover, all other constraints are independent of $\theta$, the same tuple $\{\mathbf{t}^{\prime},\boldsymbol{\mathcal{P}}^{\prime},\boldsymbol{\mathcal{A}}^{\prime}\}$
also satisfies \eqref{eq:thp_P1_c1} and \eqref{eq:basic_P1_c1}. Therefore,
$\{\mathbf{t}^{\prime},\boldsymbol{\mathcal{P}}^{\prime},\boldsymbol{\mathcal{A}}^{\prime}\}\in\mathcal{F}\left(\theta_{2}\right)$,
which establishes $\mathcal{F}\left(\theta_{1}\right)\subseteq\mathcal{F}\left(\theta_{2}\right)$.

With this result, minimizing the same objective $E$ in $\mathscr{P}1$, over a larger
feasible set cannot yield a higher optimum. Therefore, 
\[
E^*\left(\theta_{1}\right)\ge E^*\left(\theta_{2}\right),\forall\theta_{1}<\theta_{2}
\]
showing that $E^*(\theta)$ is non-increasing in $\theta$.
\end{IEEEproof}

\begin{lem}
\label{lem:monotonic_E_theta_strictly}
For any $\theta_{1}<\theta_{2}\in\{\underline{\theta},\cdots,\overline{\theta}\}$, $E\left(\theta_{1}\right)>E\left(\theta_{2}\right)$.
\end{lem}
\begin{IEEEproof}
Suppose, to the contrary, that there exist $\underline{\theta}\le\theta_{1}<\theta_{2}\le\overline{\theta}$
such that $E^*\left(\theta_{1}\right)=E^*\left(\theta_{2}\right)$. Let $\{\mathbf{t}^{\prime},\boldsymbol{\mathcal{P}}^{\prime},\boldsymbol{\mathcal{A}}^{\prime}\}\in\mathcal{F}\left(\theta_{1}\right)$ be any optimal solution of problem $\mathscr{P}2$ over $\mathcal{F}\left(\theta_{1}\right)$. Because $\mathcal{F}\left(\theta_{1}\right)\subseteq\mathcal{F}\left(\theta_{2}\right)$ due to $\theta_{1}<\theta_{2}$ and $E^*\left(\theta_{1}\right)=E^*\left(\theta_{2}\right)$ hold, $\{\mathbf{t}^{\prime},\boldsymbol{\mathcal{P}}^{\prime},\boldsymbol{\mathcal{A}}^{\prime}\}$ is also optimal for $\mathscr{P}2$ over $\mathcal{F}\left(\theta_{2}\right)$.

By the definition of $\mathcal{F}\left(\theta\right)$ in (\ref{eq:def_F_theta}), its per-$(n,t)$ \ac{rb} constraints satisfy $\sum_{k}a_{n}^{\prime}\left[k,t\right]\le\theta_{1},\forall n,t$. Hence, we have $\sum_{k}a_{n}^{\prime}\left[k,t\right]\le\theta_{1}<\theta_{2},\forall n,t.$ That is, all \ac{rb}-cap constraints are inactive (strictly slack) at the $\theta_{2}$-problem realized by $\boldsymbol{\mathcal{A}}^{\prime}$. 

From parametric sensitivity analysis (see \cite{Boy:B04}), the derivative of the optimal value with respect to a right-hand-side parameter equals the optimal Lagrange multiplier of the corresponding constraint. If the constraint is inactive (strictly slack), its multiplier is zero, and any right-hand-side relaxations do not affect the optimal value or the primal minimizers. Hence, the same solution remains optimal for all $\theta_{3}>\theta_{2}$, which implies
\[
E^*\left(\theta_{3}\right)=E^*\left(\theta_{2}\right)=E^*\left(\theta_{1}\right),\forall\theta_{3}\in\{\theta_{2}+1,\cdots,\overline{\theta}\}
\]
In particular, evaluating at $\theta_{3}=\overline{\theta}$ yields
$E^{*}=E^*(\overline{\theta})=E^*\left(\theta_{1}\right)$, which contradicts
the definition of $\overline{\theta}$ in (\ref{eq:def_theta_up}),
where $\overline{\theta}$ is the minimal $\theta$ such that $E^{*}=E^*(\theta)$. 

Therefore, the assumption is false, and we conclude that $E^*\left(\theta_{1}\right)>E^*\left(\theta_{2}\right)$
for any $\underline{\theta}\le\theta_{1}<\theta_{2}\le\overline{\theta}$.
\end{IEEEproof}

\subsection{Points on $\mathcal{C}$ are Pareto-optimal.}

For any $\theta\in\{\underline{\theta},\cdots,\overline{\theta}\}$,
suppose, to the contrary, there exists a feasible pair $(\theta^{\prime},E^{\prime})$
that weakly dominates $(\theta,E^*(\theta))$, \emph{i.e.}, 
\[
\theta^{\prime}\le\theta,E^{\prime}\le E^*\left(\theta\right),\text{ and }\left(\theta,E^*\left(\theta\right)\right)\neq\left(\theta^{\prime},E^{\prime}\right).
\]
By definition of $E^*(\cdot)$, feasibility at $\theta^{\prime}$ implies
$E^{\prime}\ge E^*\left(\theta^{\prime}\right)$. Hence 
\[
E^*\left(\theta^{\prime}\right)\le E^{\prime}\le E^*\left(\theta\right).
\]
If $\theta^{\prime}<\theta$, strictly monotonicity on $\{\underline{\theta},\cdots,\overline{\theta}\}$ according to Lemma \ref{lem:monotonic_E_theta_strictly} gives $E\left(\theta^{\prime}\right)>E\left(\theta\right)$, which contradicts $E\left(\theta^{\prime}\right)\le E\left(\theta\right)$.
If $\theta^{\prime}=\theta$, then $E\left(\theta^{\prime}\right)=E\left(\theta\right)$ force $E^{\prime}=E\left(\theta\right)$, contradicting $\left(\theta,E\left(\theta\right)\right)\neq\left(\theta^{\prime},E^{\prime}\right)$.

Thus, no feasible pair weakly dominates $(\theta,E^*(\theta))$; hence, any $(\theta,E^*(\theta))\in\mathcal{C}$ is Pareto-optimal.

\subsection{Every Pareto-optimal feasible pair lies on $\mathcal{C}$.}

Let $(\theta^{\prime},E^{\prime})$ be any feasible pair. By definition
of $E^*(\cdot)$, $E^{\prime}\ge E^*\left(\theta^{\prime}\right)$. If
$E^{\prime}>E^*\left(\theta^{\prime}\right)$, then $(\theta^{\prime},E^*\left(\theta^{\prime}\right))$
(which is feasible) strictly improves the energy objective without
worsening $\theta$, so $(\theta^{\prime},E^{\prime})$ is dominated
and cannot be Pareto-optimal. Therefore any Pareto-optimal feasible
pair must satisfy $E^{\prime}=E^*\left(\theta^{\prime}\right)$, \emph{i.e.},
it lies on $\mathcal{C}$.

\section{Proof of Corollary \ref{cor:pareto_front}\label{sec:proof_cor_pareto_front}}
Since $E^*\left(\theta\right)$ is non-increasing (strictly decreasing on $\{\underline{\theta},\cdots\overline{\theta}\}$), and $f_{1}(\cdot)$ and $f_{2}(\cdot)$ are strictly increasing, the composition $f_{2}\circ E^*\circ f_{1}$ preserves order and strictness on that interval. The mapping $\left(\theta,E\right)\mapsto(f_{1}(\theta),f_{2}(E))$ is thus order-preserving, hence the image of $\mathcal{C}$ is again a Pareto frontier.

\section{Proof of Proposition \ref{prop:opt_trans}\label{sec:proof_prop_opt_trans}}

Separate the energy consumption over status update, we have $E=\sum_{i}\sum_{n,k,t=t_{i}}^{t_{i+1}-1}a_{n}\left[k,t\right]p_{n}\left[k,t\right]$.
Then, given $\mathbf{t}$, $\mathscr{P}2$ becomes
\[
\underset{\mathbf{P} \in \mathcal{P}(\mathcal{T}),\mathbf{A}\in \mathcal{A}(\mathcal{T})}{\text{min}}\ \sum_{i}\sum_{n,k,t=t_{i}}^{t_{i+1}-1}a_{n}\left[k,t\right]p_{n}\left[k,t\right]\text{ s.t. (\ref{eq:thp_P1_c1}, \ref{eq:theta_c})}
\]
where the constraint for $\mathbf{t}$ is removed because $\mathbf{t}$ is given, which will be optimized later. 

Denote $\mathbf{\pi}_{i}=\{p_{n}\left[k,t\right],a_{n}\left[k,t\right]\}_{n\in\mathcal{N},k\in\mathcal{K},t\in\mathcal{T}_{i}}$,
where $\mathcal{T}_{i}=\{t_{i},t_{i}+1,\cdots,t_{i+1}-1\}$, it is
observed the variables for $\mathbf{x}_{i}$ and $\mathbf{\pi}_{j}$
for $i\neq j\in\mathcal{I}$ are uncoupled. Therefore, problem $\mathscr{P}2$ given $\mathbf{t}$ can be written as
\begin{align}
\sum_{i}\underset{\mathbf{\pi}_{i}}{\text{min}} & \ \sum_{n,k,t=t_{i}}^{t_{i+1}-1}a_{n}\left[k,t\right]p_{n}\left[k,t\right] \nonumber\\  
\text{s.t.} & \ \text{(\ref{eq:thp_P1_c1}) for } i, \sum_{k\in\mathcal{K}}a_{n}\left[k,t\right]\le \theta,\,\forall n\in\mathcal{N},t\in\mathcal{T}_i \nonumber\\
& \  \left\{p_m[k,t]\right\}\in \mathcal{P}(\mathcal{T}_{i}), \left\{a_m[k,t]\right\}\in \mathcal{A}(\mathcal{T}_{i})\nonumber
\end{align}

Denote the optimal value to each transmission as $E^{*}(t_{i},t_{i+1})$, then, Problem $\mathscr{P}2$ can be written as
\[
\underset{\mathbf{t}}{\text{min}}\ \sum_{i}E^{*}\left(t_{i},t_{i+1}\right)\text{ s.t. } \mathbf{t}\in \Upsilon.
\]
\section{Proof of Proposition \ref{prop:opt_P21}\label{sec:proof_prop_opt_P21}}
\emph{Path-feasible Correspondence}. Feasible $\{t_{i}\}$ in $\mathscr{P}\text{2-1}$ satisfies $1\le t_{i+1}-t_{i}\le\bar{\tau},\forall i$
and $1=t_{0}\le\cdots\le t_{I+1}=T,t_{i}\in\mathcal{T},\forall i$. By graph construction, each consecutive pair $(t_{i},t_{i+1})$ is a valid edge; hence $\{t_{i}\}$ induces a path $1\to t_{1}\to\cdots\to T_{I+1}=T+1$. Vice versa. 

\emph{Cost equivalence}. Each edge $(t_{i},t_{i+1})$ has weight $w_{i,i+1}=E^{*}(t_{i},t_{i+1})$, thus, the path length $\sum_{i}w_{i,i+1}$ equals the outer objective $\sum_{i}E^{*}(t_{i},t_{i+1})$ at the corresponding $\{t_{i}\}$. 

Accordingly, minimizing $\sum_{i}E^{*}(t_{i},t_{i+1})$ over feasible $\{t_{i}\}$ is identical to finding the shortest $1\to T+1$ in $\mathscr{G}$.

\section{Complexity Analysis for Graph $\mathscr{G}$ Construction} \label{sec:proof_complexity_G}
\begin{figure}
\begin{centering}
\includegraphics[width=1\columnwidth]{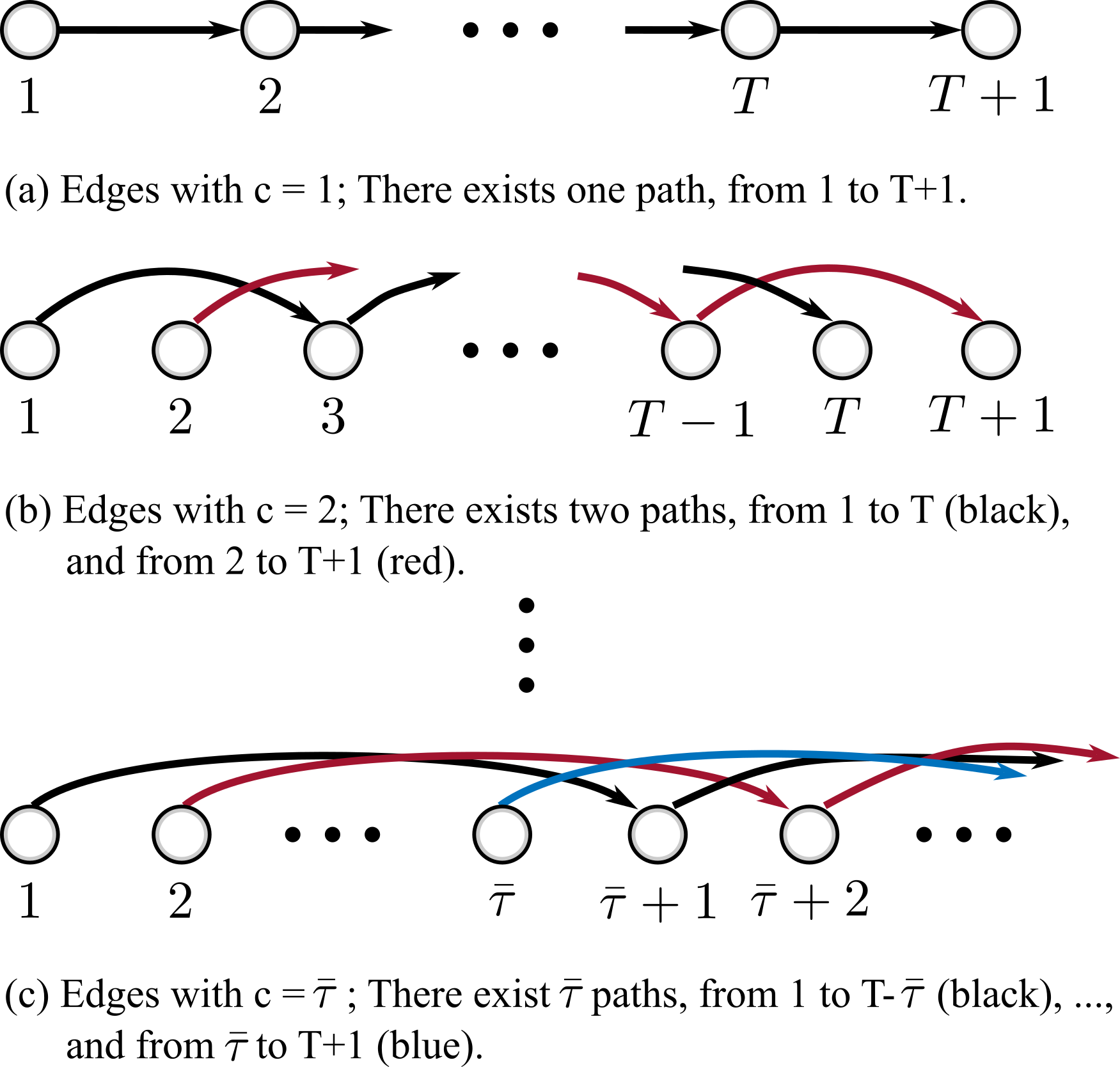}
\par\end{centering}
\caption{\label{fig:edge_illustration}Edge division according to transmission interval lengths, {\em i.e.}, $t_j-t_i$.}
\end{figure}
To analyze the computational cost of constructing $\mathscr{G}$, we partition the edges into $\bar{\tau}$ groups indexed by $c\in {1,\cdots,\bar{\tau}}$; in group $c$, every edge $(v_i,v_j)$ induces a transmission interval of length $c$, that is, $v_j-v_i=c$, as shown in Figure~\ref{fig:edge_illustration}. Since any valid edge in $\mathscr{G}$ must satisfy $1\le t_i-t_j\le \bar{\tau}$ according to the construction policy described in Section \ref{subsec:graph}, the union of all $\bar{\tau}$ groups collectively constitutes the complete edge set of $\mathscr{G}$.

For each group, there are $c$ paths from near $1$ to near $T+1$, and the cost for calculating the weight of edges in this group is less than $cT\theta\log_2(c\theta)$. For example, when $c=1$, there is only one path $1\to 2 \to 3 \to \cdots \to T+1$ and the transmission interval $1$. Therefore, the complexity is \[
\sum_i^T (t_{i+1}-t_i)\theta\log_2((t_{i+1}-t_i)\theta) =  \sum_i^T \theta\log_2(\theta) = T\log_2(\theta).
\]
When $c=2$,  there is only two paths $1\to 3 \to 5 \to \cdots \to T$ and $2\to 4 \to 6 \to \cdots \to T+1$, and the transmission interval is $2$. Therefore, the complexity is
\begin{align}
2 \sum_i^{\lfloor T/2\rfloor} (t_{i+1}-t_i)\theta\log_2((t_{i+1}-t_i)\theta) & =  2\sum_i^{\lfloor T/2\rfloor} 2\theta\log_2(2\theta) \nonumber\\ 
& \le 2T\log_2(\theta). \nonumber
\end{align}
Following the same reasoning, the computational cost for groups with $c\in {3,\cdots,\bar{\tau}}$ can be derived analogously.

As a result, the complexity for calculating all weights in the graph $\mathscr{G}$ is less than
\begin{align}
& \sum_{c=1}^{\bar{\tau}} cT\theta\log_2(c\theta) =  T\theta\sum_{c=1}^{\bar{\tau}}c\log\left(c\right)+T\theta\sum_{c=1}^{\bar{\tau}}c\log\left(\theta\right) \nonumber\\
 & =T\theta\sum_{c=1}^{\bar{\tau}}c\log\left(c\right)+T\theta\log\left(\theta\right)\sum_{c=1}^{\bar{\tau}}c \nonumber\\
 & =T\theta\sum_{c=1}^{\bar{\tau}}c\log\left(c\right)+T\theta\log\left(\theta\right)\frac{\bar{\tau}\left(\bar{\tau}+1\right)}{2} \nonumber\\
& \le T\theta\int_{c=1}^{\bar{\tau}}c\log\left(c\right)dc + T\theta\log\left(\theta\right)\frac{\bar{\tau}\left(\bar{\tau}+1\right)}{2} \nonumber\\
& = T\theta \left(\frac{\bar{\tau}^2}{2} \log_2\left(\bar{\tau}\right) -\frac{\bar{\tau}^2}{4\ln(2)}+\frac{1}{4\ln(2)} \right) \nonumber\\
& \quad\quad\quad\quad\quad\quad\quad\quad\quad\quad
+ T\theta\log\left(\theta\right)\frac{\bar{\tau}\left(\bar{\tau}+1\right)}{2}\nonumber\\
 & =\mathcal{O}\left(T\theta\bar{\tau}^{2}\log\left(\bar{\tau}\theta\right)\right) \nonumber
\end{align}
which is the order of
\begin{multline}
 \mathcal{O}\left(T\theta\bar{\tau}^{2}\log\left(\bar{\tau}\right)\right) + \mathcal{O}\left(T\theta\bar{\tau}^{2}\log\left(\theta\right)\right) \nonumber\\
 = \mathcal{O}\left(T\theta\bar{\tau}^{2}\log\left(\bar{\tau}\theta\right)\right).
\end{multline}

In summary, the complexity for constructing graph $\mathscr{G}$ is $\mathcal{O}\left(T\theta\bar{\tau}^{2}\log\left(\bar{\tau}\theta\right)\right)$.
\end{document}